\documentclass[10pt,letterpaper]{article}

\usepackage{amssymb}
\usepackage{graphicx}

\usepackage{epstopdf}

\begin{document}

\title{Black hole thermodynamics as seen through a microscopic model of a relativistic Bose gas}

\author{Jozef Sk\'akala\footnote{jozef@iisertvm.ac.in} and S. Shankaranarayanan\footnote{shanki@iisertvm.ac.in}\\
School of Physics, Indian Institute of Science,\\ 
Education and Research (IISER-TVM),\\ Trivandrum 695016, India}

\maketitle

\begin{abstract}
  Equations of gravity when projected on spacetime horizons resemble
  Navier-Stokes equation of a fluid with a specific equation of state
  \cite{Damour, Paddy1, Paddy2}.  We show that this equation of state
  describes massless ideal relativistic gas. We use these results, and
  build an explicit and simple {\it molecular} model of the fluid
  living on the Schwarzschild and Reissner-Nordstr\"om black hole
  horizons.  For the spin zero Bose gas, our model makes two
  predictions: (i) The horizon area/entropy is quantized as given by
  Bekenstein's quantization rule, (ii) The model explains the correct
  type of proportionality between horizon area and entropy. However,
  for the physically relevant range of parameters, the proportionality
  constant is never equal to $1/4$.
\end{abstract}

\section{Introduction}

There exists a popular viewpoint that general relativity, like fluid
mechanics, could be seen as a thermodynamical equilibrium, macroscopic
description of a system with completely different microscopic degrees
of freedom. The idea traces back to Sakharov \cite{Sakharov}, however,
in the last two decades there has been a surge of activities in this
direction, see, for instance,
Refs. \cite{Jacobson,Paddy4,Paddy5,Paddy0}. Similar ideas also form
the core of the analogue gravity program (see, for instance, the
review \cite{analogue}).

Three decades ago, Damour \cite{Damour} showed that the equations of
general relativity when projected onto a black-hole horizon give
Navier-Stokes equation of a 2-dimensional fluid that lives on the
black hole horizon. This allowed Damour to identify various black hole
properties with the properties of a 2-dimensional fluid. (For details
on how the various fluid characteristics arise from the spacetime
metric see the above reference.)  This approach has now been
generalized to arbitrary space-time horizons in recent works of
Padmanabhan \cite{Paddy1, Paddy2}. (For other results on fluid-gravity
correspondence see Refs. \cite{Strominger1,Strominger2,Liberati1}.)
In case the results of \cite{Damour, Paddy1, Paddy2} are not only a
curious analogy, one expects that also properties of semi-classical
black holes could be holographically obtained from statistical
mechanics of the (quantum) (2 + 1)-dimensional fluid living on the
black hole horizon.

In this work, we use results of Refs. \cite{Paddy1, Paddy2} and build
an explicit and simple {\it molecular} model of the (2+1)-dimensional
fluid. We show that the
microscopic fluid model indeed provides a holographic analogue for
different phenomena in semi-classical quantum gravity, bringing very
interesting insights into the nature of various aspects of black hole
thermodynamics. For example, the microscopic model presented here
sheds some light on the proportionality between horizon area and
entropy, and provides very interesting alternative explanation for the
Bekenstein quantization of horizon area (or entropy)
\cite{Bekenstein}. For the sake of simplicity, we consider in this
work Schwarzschild and Reissner-Nordstr\"om black holes.

There have been attempts to build microscopic models describing
physics of black hole horizons. (Microscopic models outside models of
quantum gravity, such as String Theory, or Loop Quantum Gravity.)
Specifically, in Ref. \cite{Chapline}, the authors have used
Bose-Einstein condensate as a microscopic model to describe black hole
horizon as a surface of quantum phase
transition\footnote{Bose-Einstein condensates also play important role
  in the analogue gravity models.}. In this work, we also use a model
of Bose gas, however our approach and its origins are very different
from the one in Ref. \cite{Chapline}.

In the next section, we use a molecular model to build an analogue for
the thermodynamics of 4-D Schwarzschild black-hole. In section 3, we
use a similar model to build an analogue for the thermodynamics of 4-D
Reissner-Nordstr\"om black-hole.  Finally, we end with discussion and
some suggestions for future work.

In this work we use the metric signature $(-,+,+,+)$ and set
$G=c=\hbar=k_{B}=1$.

\section{The molecular model for the thermodynamics of 
Schwarzschild  black hole}

\subsection{Model set-up}

Damour-Navier-Stokes equation is obtained by projecting Einstein's equations 
on the horizon and it identifies the fluid's pressure with the horizon 
parameters as \cite{Paddy1}
\begin{equation}\label{EQS}
p=\frac{\kappa}{8\pi}=\frac{T_{H}}{4}.
\end{equation} 
($\kappa$ is the horizon surface gravity, assumed here to be positive,
and $T_{H}$ is the Hawking temperature of the horizon.)  This result
relating fluid's pressure and horizon's surface gravity holds true for
a general null surface\footnote{For the Schwarzschild black-hole and
  in Schwarzschild radial coordinates, Damour-Navier-Stokes equation
  leads to the following simple equation
\[p_{,i}=0,\]
where $i$ labels the angular coordinates. This is a trivial way to
express the zero-th law of black hole thermodynamics, which says that
the black hole temperature (or equivalently the horizon surface
gravity) is constant across the horizon. For fluid, this means that
the pressure is constant across the fluid and also that the viscosity
terms vanish, indicating that the fluid looks like an ideal fluid.}
\cite{Paddy1} and is also shown to hold from purely thermodynamic
reasons by identifying the temporal rate of change of Einstein gravity
action with the spacetime entropy production \cite{Paddy2}. The
equation of state of the fluid (\ref{EQS}) will be the starting point
of this work.

Let us note that for the Schwarzschild black hole the above expression
can be somewhat naively derived from the definition of pressure using
the black hole entropy,
\begin{equation}\label{EQS2}
p=\frac{\partial S_{H}}{\partial A_{H}}\cdot T_{H}=\frac{T_{H}}{4}.~~~
\end{equation}
Eq. (\ref{EQS2}) boils down to the fact that for Schwarzschild black hole:
\begin{equation}\label{EQS3}
dM=p\cdot dA_{H}=p\cdot dV.
\end{equation}

In order to find the link to underlying microscopic theory, it is
natural that we are interested in stationary configurations of the
fluid. Stationary fluid configurations then naturally link to horizons
generated by null Killing field. However, there is one more essential
reason why to focus on Killing horizons and this is the fact that in
order to find some potentially interesting microscopic model, we need
to understand what is fluid's energy. The natural concept for the
energy of the fluid is the concept of Komar energy, but Komar energy
is defined only for stationary spacetimes. Therefore it is essential
that we restrict ourselves to stationary configurations and Killing
horizons. We will discuss this more in the Discussion section.

As a consequence of what was said we do the following general
assumptions that link the microscopic theory of the fluid to black
hole thermodynamics, (these assumptions are consistent with the
original paper of Damour \cite{Damour}):
\begin{itemize}
\item since the fluid is a (2+1)-dimensional fluid living on a
  horizon, the volume of the fluid is naturally the area of the
  horizon $A_{H}$,
\item the temperature of the horizon $T_{H}$ (with the chosen
  normalization of the time-like Killing field) is the temperature of
  the fluid,
\item for fluid living on a Killing horizon, the total energy of the
  fluid is the horizon Komar energy.
\end{itemize}
(Later, it will be required for the entropy of the fluid to be
identified with the black hole entropy.)

Now let us demonstrate that for a \emph{generic} Killing horizon the
equation of state (\ref{EQS}) translates into equation of state of an
ideal relativistic fluid. The Komar energy of the Killing horizon is
computed from
\begin{equation}\label{Komar0}
E=-\frac {1}{8\pi}\int_{\mathcal{H}} \xi_{a;b}d\Sigma^{ab},
\end{equation}
where $\xi$ is the suitably normalized Killing field generating the
horizon. If the surface gravity of the horizon is positive, after
short calculation one obtains
\begin{equation}\label{Komar2}
E=\frac{T_{H}V}{2}.
\end{equation}
Combining Eq.(\ref{EQS}) and Eq.(\ref{Komar2}) one gets
\[p=\frac{T_{H}}{4}=\frac{E}{2V}.\] This is an equation of state of an
2D ideal massless relativistic gas \cite{Huang}.

Our aim is to identify a microscopic model for the Schwarzschild black
hole using the fluid-gravity correspondence.  For a microscopic theory
defined by a Hamiltonian, the entropy is typically a function of three
independent parameters $S(E,N,V)$, where $E$ is system's energy, $N$
is the number of particles and $V$ is the volume. In what will follow
we utilize some aspects of the microcanonical ensemble approach. For
example the concept of temperature used in the following section will
be defined as in the microcanonical ensemble, that is, knowing
$S(E,N,V)$ we will fix $N,V$ and define $T=dE/dS$.  However, the
parameter space of Schwarzschild black hole is one-dimensional and can
be expressed by a single parameter $V=A_{H}$ (or $E$.) Therefore, if a
microscopic model is supposed to reproduce Schwarzschild black hole,
one needs to constrain the space of states of the fluid to only one
dimension. This means from the point of view of the \emph{relevant}
states of the system we will not treat the variables $E, N, V$ as
independent, but they must fulfil two constrains: $E=E(V)$ and
$N=N(V)$.

We can obtain the constrains as follows: If we interpret the fluid
energy as the Komar energy and the volume as the horizon area, they
must fulfil the relation imposed by the Schwarzschild solution:
\begin{equation}\label{volume}
V=16\pi E^{2}.
\end{equation}
This means Eq.(\ref{volume}) provides the constrain $E=E(V)$. The second 
constraint $N=N(V)$ can be derived from the other well known equation, relating the 
mass of the black-hole and Hawking temperature: 
%
\begin{equation}\label{time}
E=\frac{1}{8\pi T_{H}}.
\end{equation}
However, to derive $N(V)$ from Eq. (\ref{time}), requires information
about the microscopic model, and therefore it will be done later.  At
this stage all one needs to keep in mind is that the fluid's parameter
space is fully constrained to one dimension by Eq.(\ref{volume}) and
Eq.(\ref{time}). Furthermore, the constraints mean that for a specific
value of $V$ only a specific combination of thermodynamical potentials
$p, \mu, T$ is allowed.

In the first half of this section we have shown that the fluid is an
ideal massless relativistic gas. The relativistic gas can be either
Bose or Fermi. In the rest of this section, we show that the microscopic
model of a massless Bose gas living on a sphere predicts the form of
$N(V)$ that leads to the Bekenstein's black-hole area quantization
\cite{Bekenstein}.

\subsection{The microscopic description (from the microcanonical point of view)}

Now let us have a better look at the consequences of the ideal gas
model. The energy levels of a free non-relativistic particles living
on a sphere
 were calculated in \cite{Shabanov,Hong} as:
\begin{equation}\label{levels}
\epsilon^{nr}_{\ell}=\frac{\ell(\ell+1)+\alpha^{2}}{2R^{2}}
\end{equation} 
where $R$ is a radius of the sphere and $\alpha^{2}$ is some
constant. (In a strict sense there are differences between
\cite{Shabanov} and \cite{Hong} on whether the constant is allowed to
be non-zero. However any energy quantum spectrum can be always shifted
by a constant, which in best case is fixed by gravity considerations,
such as these, so physically we stick to the spectrum given by
Eq.(\ref{levels}) with \emph{arbitrary} $\alpha$.)  The energy levels
correspond to the Laplacian on the sphere and therefore have the
degeneracy of the spherical harmonics, given as $g_{\ell}=2\ell+1$.

As one can see from the Hamiltonian used in \cite{Shabanov}, we can
obtain spectrum of a massless relativistic scalar particle by a simple
transformation:
\begin{equation}\label{spectrum}
\epsilon^{r}_{\ell}=\sqrt{2
  \epsilon^{nr}}=\frac{\sqrt{\ell(\ell+1)+\alpha^{2}}}{\sqrt{2}\cdot
  R}=\sqrt{\frac{4\pi \{\ell(\ell+1)+\alpha^{2}\}}{V}}.
\end{equation}
This can be (through our constrains) expressed as a simple function of
temperature
\begin{equation}\label{spectrum2}
\epsilon^{r}_{\ell}=\sqrt{\{\ell(\ell+1)+\alpha^{2}\}}\cdot
T=\tilde\epsilon_{\ell}\cdot T,
\end{equation}
with
\[\tilde\epsilon_{\ell}=\sqrt{\ell(\ell+1)+\alpha^{2}}.\]
($\tilde\epsilon_{\ell}$ is independent on the black hole parameters.)
For simplicity, in this work, we utilize the spectrum
(\ref{spectrum2}) and we model the ideal relativistic massless gas by
particles with zero spin, which also means we work with Bose gas.

We have constrained our system by equations (\ref{volume}) and
(\ref{time}). One of those equations gives the constraint $E(V)$, the
other equation then implies another constraint $N(V)$. Let us now
obtain the constraint $N(V)$: Note that in a microcanonical ensemble
when $T$ is much larger than spacing between the quantum energy
levels, the mean particle's energy linearly decreases with
temperature, as prescribed by the equipartition law\footnote{The
  equipartition law was used in the context of Schwarzschild black
  hole in \cite{Paddy3}.}. It can be shown for a simple harmonic
oscillator\footnote{The spacing between the energy levels from
  Eq.(\ref{spectrum}) approaches for large $\ell$ harmonic
  oscillator's energy levels, for small energy levels the difference
  between the levels is larger than in the case of oscillator.} that
the equipartition law ceases to hold for very low temperatures, when
the temperature is of a comparable value to the discrete spacing
between the energy levels. Furthermore, for temperatures of a value
comparable to the spacing between the energy levels the average
particle's energy is very close to the ground state energy. From
Eq.(\ref{spectrum2}) one observes that, in our constrained system, the
temperature always is of a comparable value to the spacing between the
energy levels, which means the equipartition law is not applicable.
 
This in turn implies that our free gas is supposed to have mean
particle's energy very near the ground state. In such case we can use
the spectrum (\ref{spectrum2}) to see that the mean particle's energy
must be:
\[\bar{E}=\gamma\cdot T,\]
where $\gamma\geq |\alpha|$ and $\gamma$ being approximately of the
same order as $|\alpha|$. Since
\[E=N\bar{E}=N\gamma\cdot T\equiv (8\pi T)^{-1},\]
this implies
\[N=\frac{1}{8\pi\gamma T^{2}}=\frac{A_{H}}{2\gamma}.\]
Therefore 
\begin{equation}\label{Bekenstein}
A_{H}=2\gamma\cdot N.
\end{equation}
Eq.(\ref{Bekenstein}) is the desired second constraint $N(V)$. This is
one of the main results of this work, and we would like to stress the
following points:

\begin{itemize}
\item The derivation of $N(V)$ is based on few insights and
  presuppositions in the microcanonical ensemble approach. In the
  following section, we derive the same result more rigorously. In
  particular, in the next section, we will use the result given by
  Eq.(\ref{Bekenstein}) as an Ansatz and it will be shown that the
  Ansatz is correct in the sense that it fulfils all the constrains
  of the model. This means the relation (\ref{Bekenstein}) is a
  \emph{direct consequence} and \emph{prediction} of our microscopic
  model.

\item Eq.(\ref{Bekenstein}) is remarkable as it gives Bekenstein's
  \cite{Bekenstein} quantization of the black hole horizon area (since
  $N$ is by definition a natural number). It gives a completely new
  and independent insight into Bekenstein's result. Furthermore, the
  insight does not rely on the quantum theory, only on the fluid
  interpretation of gravity. (The constant $\gamma$ is here arbitrary,
  but can be fixed to obtain the most popular form of Bekenstein type
  of spectrum as $\gamma=4\pi$. Since it will be shown that
  $\gamma=|\alpha|$, it can be demonstrated that this fixing gives
  wave-length of the particle in the ground state equal to the
  circumference of the black hole horizon.) Note that the linear
  proportionality between the horizon area and the number of degrees
  of freedom is a starting assumption in the grand-canonical
  statistical analysis of Schwarzschild black hole in
  Ref. \cite{Gour}.
\item The formula (\ref{Bekenstein}) has also a deeper meaning; it is
  problematic to speak about density of fluid's degrees of freedom, as
  to speak about density one needs pre-defined geometry. However,
  geometry is in this view a macroscopic property constituted by the
  fluid. In some sense it is natural to fix the way how the geometry
  is constituted by the fluid by relating it to the density of the
  fluid's degrees of freedom as:
\[\frac{N}{A_{H}}=(2\gamma)^{-1}=const. ~.\]

This also explains why in the fluid-gravity approach one obtains
analogue of Navier-Stokes equation \cite{Paddy1} (and energy diffusion
equation given by the Raychhadhuri equation), but not the (rest-)mass
conservation equation. The fact that the particles are massless, and
the density of particles is always the same constant, means that mass
conservation equation is fulfilled by a simple identity (its
information rests only in how geometry links to the fluid particles).

\item One of the standard problems with black hole physics is the fact
  that black holes have usually negative heat capacity. This means for black
  holes one cannot define canonical partition function \cite{Paddy6}
  and it seems black holes cannot be represented by well known
  microscopic models of a fluid. The key observation is that our model
  is very different: By fixing volume and the number of particles, and
  computing derivative of $E(T)$ one obtains positive heat
  capacity. However since our model is one-dimensional constrained
  model in which change in temperature leads \emph{simultaneously} to
  change in all fluid parameters (energy, number of particles and
  volume), the derivative of $E(T)$ does not correspond to a standard
  notion of heat capacity and can be negative.

\end{itemize}

\subsection{More rigorous statistical ensemble calculation }

One can confirm the insights from the last section through a more
rigorous statistical ensemble calculation of our constrained
system. The calculation is a hybrid between canonical and
grand-canonical ensemble, therefore we refer to it as
\emph{statistical ensemble calculation}. The number of particles is
not kept fixed (which makes it look like grand-canonical ensemble),
but the system is constrained to only one parameter (which we choose
to be the energy), which makes it look as a canonical ensemble. We
want to stress that our calculation is still semi-classical and
therefore we assume our system to be in a sufficiently low temperature
regime. (This is the regime of sufficiently large black holes where
semi-classical description is reasonably close to the reality.) More
precisely, the sufficiently low temperature regime means that
$T/T_{p}<<1$, where $T_{p}$ is Planck temperature. Since we use Planck
units, $T_{p}=1$, and therefore in Planck units the condition of
semi-classicality translates to $T<<1$.

Let us start from the first principles and assume our system to be in
a contact with a huge reservoir. Then the probability of a state with
energy $E$ for a Bose system is given by:

\begin{eqnarray}\label{prob}
p(E) \sim\Omega\{n_{\ell}\}\exp\left\{-\frac{d S(E,V(E),N(E))}{d E}\cdot
E\right\}=~~~~~~~~~~~~~~~~~~~~~~~~~~~~~~~~~~~~~~~~~~~~~~~~~\nonumber\\ \Omega\{n_{\ell}\}\exp\left\{-\frac{\partial
  S(E,V(E),N(E))}{\partial E}\cdot E-\frac{\partial
  S(E,V(E),N(E))}{\partial V}\frac{\partial V}{\partial E}\cdot
E-\right.~~~~~\\
\left.\frac{\partial S(E,V(E),N(E))}{\partial N}\frac{\partial N}{\partial
  E}\cdot
E\right\}=\Omega\{n_{\ell}\}\exp\left\{-\frac{E}{T}+ f(T) \, \frac{\partial
  N}{\partial E}\cdot
E\right\}=~~\nonumber\\
\Omega\{n_{\ell}\}\exp\left\{-\frac{E}{T}+ f(T) \, \frac{16\pi}{\gamma}\cdot
E^{2}\right\}.~~~~~~~~~~~\nonumber
\end{eqnarray}

Here
\begin{equation}
\Omega\{n_{\ell}\}=\prod_{\ell}\frac{(n_{\ell}+g_{\ell}-1)!}{n_{\ell}!(g_{\ell}-1)!} \quad ~~\mbox{and} ~~\quad
 f(T)=\frac{\mu-2\gamma p}{T}, 
\end{equation}

with $g_{\ell}$ being the degeneracy of the energy level and $\mu$ chemical potential. The
variables in Eq.(\ref{prob}) are subject to two constrains
\begin{equation}\label{const1}
\sum_{\ell}\epsilon_{\ell}n_{\ell}-E=0,
\end{equation}
and
\begin{equation}\label{const2}
\sum_{\ell}n_{\ell}-N=\sum_{\ell}n_{\ell}-\frac{8\pi E^{2}}{\gamma}=0.
\end{equation}
In constraint (\ref{const2}) we substituted for $N$ a function of $E$
using Eq.(\ref{Bekenstein}) and Eq.(\ref{volume}).

For $n_{\ell}>>1$ one can use Stirling's formula and transform Eq.(\ref{prob}) to:
\begin{eqnarray}\label{Bose2}
p(E)\sim \exp\left(-\frac{E}{T}+\frac{16\pi f(T)E^{2}}{\gamma}+\right.~~~~~~~~~~~~~~~~~~~~~~~~~~~~~~~~~~~~~~~~~~~~~~\nonumber\\
+\left.\sum_{\ell}\left[(n_{\ell}+g_{\ell})\{\ln(n_{\ell}+g_{\ell})-1\}-n_{\ell}\{\ln(n_{\ell})-1\}-\ln([g_{\ell}-1]!)\right]\right).
\end{eqnarray}

If we assume that the value of $\bar{E}(T)$ is given by the extremum
of $p(E)$, then we obtain using Lagrange multipliers the following
equations
\[\frac{\partial}{\partial n_{\ell'}}:~~~~\ln\left(\frac{n_{\ell'}+g_{\ell'}}{n_{\ell'}}\right)+\tilde\lambda\cdot\epsilon_{\ell'}+\lambda=0,\]
\[\frac{\partial}{\partial E}:~~~~-\frac{1}{T}+\frac{32\pi f(T) E}{\gamma}-\tilde\lambda-\frac{16\pi E\lambda}{\gamma}=0,\]
\[\frac{\partial}{\partial\tilde\lambda}:~~~~\sum_{\ell}\epsilon_{\ell}n_{\ell}-E=0,\]
\[\frac{\partial}{\partial\lambda}:~~~~\sum_{\ell}n_{\ell}-\frac{8\pi E^{2}}{\gamma}=0.\]
Here $\tilde\lambda$ and $\lambda$ are Lagrange multipliers
corresponding to the two constraints (\ref{const1}) and
(\ref{const2}). Take the first equation and substitute for
$\tilde\lambda$ from the second equation to obtain:
\begin{equation}\label{Ext1}
\bar{n}_{\ell'}=\frac{g_{\ell'}}{\exp\left(\left[\frac{16\pi E
      \lambda}{\gamma}+\frac{1}{T}- \frac{32\pi
      f(T)E}{\gamma}\right]\epsilon_{\ell'}-\lambda\right)-1}.
\end{equation}
Let us further use the fact that $E$ must be a function of temperature
given by Eq.(\ref{time}).  Then equation (\ref{Ext1}) can be rewritten
as
\begin{equation}\label{Ext2}
\bar{n}_{\ell'}=\frac{g_{\ell'}}{\exp\left(\left[\frac{2\lambda}{\gamma}+1-
    \frac{4
      f(T)}{\gamma}\right]\tilde\epsilon_{\ell'}-\lambda\right)-1}.
\end{equation}
($\tilde\epsilon_{\ell'}$ was defined before and is independent on
external parameters.)

Before we proceed, it may be worth pointing out that the equation
(\ref{Ext2}) has certain limitations.
The limitations arise due to the use of Stirling's approximation and
the fact that in general for higher occupation numbers (where the
numbers are small) the above formula could give inaccurate results.
Despite some quantitative modifications (due to changes in the
asymptotic behavior of $\bar{n}_{\ell}$), in most of the cases one can
still expect that the exact results go qualitatively along similar
lines as the ones derived via Eq.(\ref{Ext2}). For example constrains
(\ref{const1}) and (\ref{const2}) could fix (slightly different)
$\gamma(\alpha)$ and $f(T)$, however, the final conclusions can be
expected to remain unchanged.

A necessary condition for every occupation number is that it is
positive for each level. This leads to the condition
\begin{equation}\label{const3}
\frac{2\lambda}{\gamma}+1- \frac{4 f(T)}{\gamma}\geq 0~~~~~~\forall \,\, T.
\end{equation}
Now let us explore the possibilities of obtaining divergent number of
particles for the asymptotic case $T=0$. (This is a necessary
condition that our model must fulfil and means that in the limiting
case of zero temperature it corresponds to a black hole with infinite
horizon area and infinite mass.) One possibility would be that
\[\frac{2\lambda}{\gamma}+1- \frac{4 f(0)}{\gamma}= 0,\]
and $\lambda\leq 0$, which would lead to a constant finite non-zero
distribution of particles over states. This, however, would contradict
our assumption that the average particle's energy is close to the
ground state.

The only remaining possibility (how to obtain the infinite number of
particles) is if the infinite number of particles are specifically in
the ground state. This happens if:
\[\lambda=\frac{(4f(0)-\gamma)|\alpha|}{2|\alpha|-\gamma}.\]
One can then easily derive from Eq.(\ref{Ext2}) that:
\begin{equation}\label{ratio}
0<\sum_{\ell=1}^{\infty}\bar{n}_{\ell} <\infty
~~~\hbox{as}~~~ T=0.
\end{equation}
where $n_{0}$ would be the occupation number of the ground state
$\ell=0$.  Since also
$\sum_{\ell=1}^{\infty}\tilde\epsilon_{\ell}\bar{n}_{\ell}$ (the sum
of energies above the ground state energy) clearly converges, this
means $\gamma=|\alpha|$ and
\begin{equation}\label{lambda}
\lambda=4f(0)-|\alpha|.
\end{equation}

It is important to realize that such a solution, represented by
Eq.(\ref{lambda}), can be always chosen for $\lambda$. This is because
the only consistency conditions are the two constrains (\ref{const1})
and (\ref{const2}), in which constraint (\ref{const1}) only leads for
$T<<1$ to our condition $\gamma=|\alpha|$ and the constraint
(\ref{const2}) can always be fulfilled, since it just determines the
correct choice of $f(T)$, which means the correct choice of the
chemical potential.

Eq.(\ref{lambda}) together with Eq.(\ref{const3}) gives the following
condition on $f(0)$:
\begin{equation}\label{lowbound}
f(0)>\frac{\gamma}{4}=\frac{|\alpha|}{4}.
\end{equation}
Eq.(\ref{Ext2}) can be further rewritten as
\begin{equation}
\bar{n}_{\ell'}=\frac{g_{\ell'}}{\exp\left(\left[\frac{8
      f(0)}{|\alpha|}-1- \frac{4
      f(T)}{|\alpha|}\right]\tilde\epsilon_{\ell'}-4f(0)+|\alpha|\right)-1},
\end{equation}
which gives in the $T=0$ asymptotic case
\begin{equation}\label{Ext3}
  \bar{n}_{\ell'}=\frac{g_{\ell'}}{\exp\left(\left[-1+ \frac{4
          f(0)}{|\alpha|}\right]\tilde\epsilon_{\ell'}-4f(0)+|\alpha|\right)-1}.
\end{equation}

Now with these observations one can fix the function $f(T)$ through
Eq.(\ref{const2}). If $T<<1$ the following equation is satisfied:
\[N\approx\frac{1}{\exp\{-4\cdot ( f(T)-f(0))\}-1},\]
(the ground state is non-degenerate).
This fixes 
\[f(T)=f(0)-\frac{1}{4}\ln\left(1+\frac{1}{N}\right)=
f(0)-\frac{1}{4}\ln\left(1+\frac{|\alpha|\cdot T^{2}}{2\pi}\right).\]
Therefore $[f(T)-f(0)]\in O(T^{2}).$

\subsection{Calculating the entropy}

There is one significant consistency check and this is the fact that
we require the fluid's entropy to match the black hole's entropy. As
previously mentioned, in the low-temperature limit we need to match
the fluid entropy with the semi-classical black-hole thermodynamics
results.

Let us use the first law
of thermodynamics:
\begin{equation}\label{1law}
TdS=dE+p\cdot dV-\mu\cdot dN,
\end{equation}
which leads to ($T$ small again)
\[dS=\frac{dE}{T}-f(T)\cdot dN\approx \frac{dE}{T} -\frac{2f(0)\cdot
  dE}{|\alpha|T}.\] This automatically gives entropy to be
proportional to the horizon area $\sim A_{H}$. However, claiming that
entropy must be a growing function of $A_{H}$ gives the following
upper bound on $f(0)$:
\[f(0)<\frac{|\alpha|}{2}.\] Since the positivity of occupation
numbers implies $f(0)>|\alpha|/4$, it means that $f(0)\in
(|\alpha|/4,|\alpha|/2)$. This means the right proportionality given
by $\sim 1/4$ cannot be reached as it would require
$f(0)=0$. Therefore, our microscopic fluid model predicts that
\emph{only part} of the black hole entropy can be explained by the
fluid. All this means is that our model indeed leads to the correct
linear proportionality between entropy and the horizon area, but with
a lower factor than $1/4$. (The proportionality factor depends on the
free parameter $f(0)$ and lies in the interval $(0,1/8)$.) In Appendix
\ref{A} we suggest that similar entropy calculation excludes fermionic
model.

One would possibly like to derive the same result in a more ``proper''
manner, from the statistical ensemble calculation. However, it is
non-trivial to determine from the partition function what is the
entropy of our system for $T<<1$. It is true that the degeneracy of
the ground state, in which dominant number of particles reside for
$T=0$, is zero, however there are finite occupation numbers when $T=0$
for the higher energy levels. These occupation numbers contribute to
non-zero entropy. To calculate the entropy (e.g if it is convergent,
or divergent at $T=0$) from the statistical approach based on
Eq.(\ref{Bose2}) seems to be impossible, as the approximations (e.g
Stirling formula) used in this section fail for the higher level
occupation numbers. Despite of this, let us make a qualitative
observation that as $T$ approaches zero the higher level occupation
numbers grow (and approximate a finite constant value) and therefore
entropy grows as $T\to 0$. This is consistent with what one observes
for Schwarzschild black hole.

In the next section, we repeat the above analysis for
Reissner-Nordstr\"om black-hole and show that we get the correct
proportionality constant for certain range of $|Q|/M$.

\section{The molecular model for the thermodynamics of
  Reissner-Nordstr\"om black hole}

\subsection{Model set-up}

Reissner-Nordstr\"om black hole is a well known spherically symmetric
static solution of Einstein equations in electrovacuum. The spacetime
is labelled by two diffeomorphism invariant parameters $M,Q$, where
$Q$ is the electromagnetic charge. For $0<|Q|/M <1$ the idealized
mathematical solution has two horizons, the outer black hole, and the
inner, Cauchy horizon. Following the idea that thermodynamical properties
of horizons are universal \cite{Paddy7, Skakala} we will consider
thermodynamics of both the horizons in question. (Since the
interest of this work are principal questions on how idealized thermodynamical properties of horizons could be explained by thermodynamical properties of fluid, we will take the inner horizon for this purpose seriously and omit
the discussion of its physical relevance and instability.)

Let us denote the two horizons with the $\pm$ notation, where the
symbol ``$+$'' symbolizes the outer and the symbol ``$-$'' symbolizes
the inner horizon. In this section, we will derive the fluid microscopic
model for both of the horizons simultaneously, however one has to keep
in mind that the model of the fluid living at the outer horizon and
the model of the fluid living at the inner horizon are two separate
models.

Let us repeat and slightly update the connections between fluid and
black hole properties (suggested already by original Damour paper
\cite{Damour}) that were made in the previous section:

\begin{itemize}

\item The fluid's energy is the Komar energy of the horizon.

\item The fluid's temperature is the temperature associated with the
  horizon.

\item The fluid's volume is the area of the horizon.  Since the
  horizon area will represent the 2D volume of the fluid, we denote it
  by the letter(s) $V_{\pm}$, rather than using the usual notation.

\item The black hole charge is the charge of the fluid.

\end{itemize}

Again, the equation of state for free relativistic gas, that was
derived in previous section for horizons with positive surface gravity
can be demonstrated to hold true for fluid living on each of the
horizons. (In the Reissner-Nordstr\"om case the assumption that
surface gravity is positive no longer holds true for the inner
horizon, but as we will further demonstrate, negative surface gravity
still gives the same fluid's equation of state.) Taking the horizon
temperatures to be
\[T_{\pm}=\frac{|\kappa_{\pm}|}{2\pi},\] where $\kappa_{\pm}$ are the
surface gravities, the equation of state for the fluid on the horizon
turns to be
\begin{equation}\label{E1}
p_{\pm}=\pm \frac{T_{\pm}}{4}.
\end{equation}

The Komar energies calculated from the usual formula (\ref{Komar0})
are given by Smarr-type of formula:
\begin{equation}\label{E2}
E_{\pm}=\pm\frac{T_{\pm}V_{\pm}}{2}.
\end{equation}
Then the equation of state (\ref{E1}) together with (\ref{E2}) implies
for both horizons the same result as in the previous section:
\begin{equation}\label{E3}
p_{\pm}=\frac{E_{\pm}V_{\pm}}{2}.
\end{equation}
This means the equation of state (\ref{E3}) holds generally for
Einstein gravity, irrespective of whether horizon's surface gravity is
positive, or negative.

Let us end this part by introducing few useful relations. 
The horizon areas can be expressed as
\begin{equation}
V_{\pm}=4\pi\cdot (M\pm\sqrt{M^{2}-Q^{2}})^{2},
\end{equation}
from which one can derive the following: 
\begin{equation}
M=\sqrt{\frac{\pi}{V_{\pm}}}\cdot\left(\frac{V_{\pm}}{4\pi}+Q^{2}\right).
\end{equation}
Further one can calculate
\begin{equation}
\sqrt{M^{2}-Q^{2}}=\pm\frac{1}{4\pi}\sqrt{\frac{\pi}{V_{\pm}}}\cdot(V_{\pm}-V_{ext}),
\end{equation}
with $V_{ext}$ being horizon area in the extremal case ($|Q| = M$), therefore
$V_{ext}=4\pi Q^{2}$.  
The horizon temperatures are given by:
\begin{equation}\label{temperatureRN}
T_{\pm}=\frac{2\sqrt{M^{2}-Q^{2}}}{V_{\pm}}=\pm\frac{1}{2\pi}\sqrt{\frac{\pi}{V_{\pm}}}\left(1-\frac{V_{ext}}{V_{\pm}}\right),~~~
\end{equation}
and Komar energies $E_{\pm}$ are given by:

\begin{equation}\label{KomarRN}
E_{\pm}=\pm\frac{V_{\pm}T_{\pm}}{2}=\pm\sqrt{M^{2}-Q ^{2}}=\frac{1}{4\pi}\sqrt{\frac{\pi}{V_{\pm}}}\cdot(V_{\pm}-V_{ext}).
\end{equation}

\subsection{Microscopic description (from the microcanonical point of view)}

Let us again consider that the energy spectrum of relativistic ideal
gas living on a sphere is given by Eq.(\ref{spectrum}), with some
ground state energies $\alpha_{\pm}$.  Now repeat the arguments from
the Schwarzschild case: The energy spectrum spacing behaves for
\emph{fixed EM charge} as $\sim V^{-1/2}$ and from
(\ref{temperatureRN}) we can again observe that the temperature is
always comparable to the energy level spacing. This means that
equipartition theorem does not apply. This is in fact good news, as
the Komar energy of the inner horizon is always negative, which cannot
be explained by equipartition theorem, but can be explained if the
ground state energy of the gas on the inner horizon has a negative
value. (This means $\alpha_{-}<0$.)

Analogous to the Schwarzschild case, let us consider that the state of
microcanonical ensemble for each horizon is described by four
parameters $E_{\pm}$, $Q$, $N_{\pm}$, $V_{\pm}$. However, since we know that
Reissner-Nordstr\"om black hole is only a two parametric model, we
will need two constraints that reduce the dimension of the space of
fluid states to two dimensions. Both of these constraints will be
obtained in a complete analogy to the Schwarzschild case from the
previous section. The first constraint is in fact the equation
(\ref{KomarRN}). For the second constraint one can use the equation
(\ref{temperatureRN}). However, as in the previous Schwarzschild case,
we would like to obtain constraints as translated to the language of
$E_{\pm}$, $Q$, $N_{\pm}$, $V_{\pm}$. We use again insight telling us that for the
relevant values of temperatures $T_{\pm}$:
\begin{equation}
E_{\pm}=\pm \frac{\tilde\gamma_{\pm} (Q)\cdot N_{\pm}}{\sqrt{V_{\pm}}},~~~~N_{\pm}\in\mathbb{N}
\end{equation} 
where $\tilde\gamma_{\pm} (Q)>0$. Eq. (\ref{KomarRN}) implies:
\begin{equation}\label{BekensteinRN}
2\gamma_{\pm} (Q)N_{\pm}=\pm(V_{\pm}-V_{ext})=|V_{\pm}-V_{ext}|,
\end{equation}
with $\gamma_{\pm} (Q)=2\sqrt{\pi}\cdot\tilde\gamma_{\pm} (Q)$. 
One can also express $N_{\pm}$ more conveniently through $E_{\pm}$ as:
\begin{equation}
N_{\pm}=\pm\frac{4\pi}{\gamma_{\pm}}\left(E_{\pm}^{2}+E_{\pm}\sqrt{E_{\pm}^{2}+Q^{2}}\right).
\end{equation}

Eq.(\ref{BekensteinRN}) is an interesting generalization of
Schwarzschild result showing that for a fixed charge ($Q$), the
spectrum distancing the outer/inner horizon area from the extremality
is equispaced. (Similar results were obtained via very different
methods also in \cite{Kunstatter1, Kunstatter2}.) Also, as intuitively
expected, the non-statistical limit $N_{\pm}\to 0$ corresponds to
extremality.

Let us briefly discuss $\gamma_{\pm}(Q)$ function and argue that it is
in fact constant (independent on $Q$). The horizon is an equipotential
surface and the charged two dimensional gas in an equilibrium state has on
average zero potential energy. (This allows to model
the equilibrium state as a free gas.)  As a consequence, further in
the calculations we assume that $\gamma_{\pm}$ is a constant and this
leads to a natural Bekenstein-type generalization of the horizon area
spectrum from the Schwarzschild case. (This is because the spacing in
the area spectrum can be in such case, for $\gamma_{+}=\gamma_{-}$, a
``universal'' quantity.) The spectrum (\ref{BekensteinRN}) is to some
extend an Ansatz, but again, in analogy to Schwarzschild case it will
be shown later in the section that it
satisfies our microscopic model. Therefore it is indicated by this
paper that, as shown more times before (see for example \cite{Skakala}),
Bekenstein spectra for horizon area/entropy are a robust result. (In
this case they also hold true for the inner Reissner-Nordstr\"om black
hole horizon.)

\subsection{More rigorous statistical calculation }

We continue with generalization of the calculations from the previous
section. We keep here the number of particles non-fixed, however the
system is still subject to two constraints (\ref{KomarRN}) and
(\ref{BekensteinRN}).  As before, the probability distribution over
states with energies $E_{\pm}$ is:
\begin{eqnarray}\label{probRN}
p(E_{\pm}) \sim\Omega\{n_{\ell}\}\exp\left\{-\left(\frac{\partial S_{\pm}(E_{\pm},Q,V_{\pm}(E_{\pm},Q),N_{\pm}(E_{\pm},Q))}{\partial E_{\pm}}\cdot E_{\pm} +\right.\right.~~\nonumber\\
\left.\left.+\frac{\partial S_{\pm}(E_{\pm},Q,V_{\pm}(E_{\pm},Q),N_{\pm}(E_{\pm},Q))}{\partial Q}\cdot Q\right)\right\} =~~~~~~~\\
=\Omega\{n_{\ell}\}\exp\left\{-\frac{E_{\pm}}{T_{\pm}}+ f_{\pm}(T_{\pm},\phi_{\pm}) \, \left(\frac{\partial
  N_{\pm}}{\partial E_{\pm}}\cdot
E_{\pm}+\frac{\partial N_{\pm}}{\partial Q}\cdot Q\right)+\frac{\phi_{\pm}}{T_{\pm}}\cdot Q\right\}.~\nonumber
\end{eqnarray}

Here, again,
\begin{equation}
\Omega\{n_{\ell}\}=\prod_{\ell}\frac{(n_{\ell}+g_{\ell}-1)!}{n_{\ell}!(g_{\ell}-1)!}
\end{equation}
and
\begin{equation}
 f_{\pm}(T_{\pm},\phi_{\pm})=\frac{\mu_{\pm}-2\gamma_{\pm} p_{\pm}}{T_{\pm}}, 
\end{equation}
with $g_{\ell}$ being the degeneracy of the energy level and $\mu$
chemical potential. The thermodynamical potential related to $Q$ is
\[\phi_{\pm}=-T_{\pm}\frac{\partial S_{\pm}}{\partial Q},\]
which for the Reissner-Nordstr\"om solution and the outer horizon
gives the electrostatic potential value at the outer horizon (at the
inner horizon it gives the potential with the opposite sign).  We also
omitted the $\pm$ subscripts in case of $n_{\ell}$, (for keeping the
notation less complicated), but they are implicitly assumed. The
variables in Eq.(\ref{probRN}) are subject to two constrains
\begin{equation}\label{const1RN}
\sum_{\ell}\epsilon_{\ell}n_{\ell}-E_{\pm}=0,
\end{equation}
and
\begin{equation}\label{const2RN}
\sum_{\ell}n_{\ell}-N_{\pm}=\sum_{\ell}n_{\ell}\mp\frac{4\pi}{\gamma_{\pm}}\left(E_{\pm}^{2}+E_{\pm}\sqrt{E_{\pm}^{2}+Q^{2}}\right)=0.
\end{equation}

For $n_{\ell}>>1$ one can use Stirling's formula and transform
Eq.(\ref{prob}) to:
\begin{eqnarray}\label{Bose2RN}
p(E_{\pm})\sim \exp\left(-\frac{E_{\pm}-\phi_{\pm}Q}{T_{\pm}}+f_{\pm}(T_{\pm},\phi_{\pm})\left(E_{\pm}\frac{\partial N_{\pm}}{\partial E_{\pm}}+Q\frac{\partial N_{\pm}}{\partial Q}\right)+\right.~~~~~~~~~~~~~~\nonumber\\
\left.\sum_{\ell}\left[(n_{\ell}+g_{\ell})\{\ln(n_{\ell}+g_{\ell})-1\}-n_{\ell}\{\ln(n_{\ell})-1\}-\ln([g_{\ell}-1]!)\right]\right).~~
\end{eqnarray}
If we assume that the value of $\bar{E_{\pm}}(T_{\pm})$ is given by
the extremum of $p(E_{\pm})$, then we obtain using Lagrange
multipliers the following equations
\[\frac{\partial}{\partial n_{\ell'}}:~~~~\ln\left(\frac{n_{\ell'}+g_{\ell'}}{n_{\ell'}}\right)+\tilde\lambda\cdot\epsilon_{\ell'}+\lambda=0,\]
\[\frac{\partial}{\partial E_{\pm}}:~~~~-\frac{1}{T_{\pm}}+f_{\pm}(T_{\pm},\phi_{\pm})\frac{\partial}{\partial E_{\pm}}\left(\frac{\partial N_{\pm}}{\partial E_{\pm}}E_{\pm}+\frac{\partial N_{\pm}}{\partial Q}Q\right)-\tilde\lambda_{\pm}-\frac{\partial N_{\pm}}{\partial E_{\pm}}\lambda_{\pm}=0,\]
\[\frac{\partial}{\partial\tilde\lambda_{\pm}}:~~~~\sum_{\ell}\epsilon_{\ell}n_{\ell}-E=0,\]
\[\frac{\partial}{\partial\lambda_{\pm}}:~~~~\sum_{\ell}n_{\ell}\mp\frac{4\pi}{\gamma_{\pm}}\left(E_{\pm}^{2}+E_{\pm}\sqrt{E_{\pm}^{2}+Q^{2}}\right)=0,\]
\[\frac{\partial}{\partial Q}:~~~ \frac{\phi_{\pm}}{T_{\pm}}+f_{\pm}(T_{\pm},\phi_{\pm})\frac{\partial}{\partial Q}\left(\frac{\partial N_{\pm}}{\partial E_{\pm}}E_{\pm}+\frac{\partial N_{\pm}}{\partial Q}Q\right)-\lambda_{\pm}\cdot\frac{\partial N_{\pm}}{\partial Q}=0.\]
Here $\tilde\lambda_{\pm}$ and $\lambda_{\pm}$ are Lagrange
multipliers corresponding to the two constraints (\ref{const1RN}) and
(\ref{const2RN}).

From the last equation one obtains
\begin{equation}\label{lambdaRN}
\lambda_{\pm}=\pm \frac{\gamma}{4\pi}\frac{\phi_{\pm}}{T_{\pm}}\frac{\sqrt{E_{\pm}^{2}+Q^{2}}}{E_{\pm}Q}+2 f_{\pm}(T_{\pm},\phi_{\pm}),~~~~~Q\neq 0.
\end{equation}
Similar to the Schwarzschild case, we get, 
\begin{eqnarray}
\bar{n}_{\ell'}=\frac{g_{\ell'}}{\exp\left(\left[\frac{[\lambda_{\pm}-f_{\pm}(T_{\pm}\phi_{\pm})]}{\sqrt{V_{\pm}}}\cdot\frac{\partial N_{\pm}}{\partial E_{\pm}}+\frac{1}{T_{\pm}\sqrt{V_{\pm}}}-
    \frac{
      f_{\pm}(T_{\pm},\phi_{\pm})}{\sqrt{V_{\pm}}}\left(\frac{\partial^{2} N_{\pm}}{\partial E_{\pm}^{2}}E_{\pm}+\frac{\partial^{2}N_{\pm}}{\partial E_{\pm}\partial Q}Q\right)\right]\tilde\epsilon_{\ell'}-\lambda_{\pm}\right)-1}.\nonumber
\end{eqnarray}
($\tilde\epsilon_{\ell'}$ is defined as in previous section:
$\tilde\epsilon_{\ell'}=\sqrt{\ell'(\ell'+1)}\pm \alpha_{\pm}$.)  Note
that $\lambda_{\pm}$ can be rewritten through another free parameter
$\phi$ via Eq.(\ref{lambdaRN}).

Let us now explore the general limit $Q^{2}/M^{2}\to K$ (where $0\leq
K\leq 1$) and $T_{+}<< 1$. Then within this limit, the following approximation 
is valid:
\[E_{\pm}\approx \pm M\sqrt{1-K}.\]
It can be further shown that in this limit ($T_{+} << 1$)
\begin{itemize}
\item $\frac{1}{T_{\pm}\sqrt{V_{\pm}}}\to C_{1\pm}\neq 0$,
\item $\frac{\partial N_{\pm}}{\partial E_{\pm}}\frac{1}{\sqrt{V_{\pm}}}\to C_{2\pm}\neq 0,$
\item $\left(E_{\pm}\frac{\partial^{2}N_{\pm}}{\partial E_{\pm}^{2}}+Q\frac{\partial^{2} N_{\pm}}{\partial E_{\pm}\partial Q}\right)\frac{1}{\sqrt{V_{\pm}}}\to C_{3\pm}\neq 0.$
\end{itemize}
($C_{1\pm}$, $C_{2\pm}$ and $C_{3\pm}$ are some constants defined by the limit and they will be later explicitly calculated.) If one again imposes that in the limiting case 
$T_{+}=0$, and $\phi_{\pm}=\phi_{0\pm}$ one will end up with infinite number of particles in the ground state,
one obtains:
\[ \lambda_{\pm}=\frac{(f_{\pm}(0,\phi_{0\pm}) \{C_{2\pm}+C_{3\pm}\} -
  C_{1\pm})\alpha_{\pm}}{C_{2\pm}\alpha_{\pm}\mp 1}. \] 
Parameter $\alpha_{\pm}$ comes from Eq.(\ref{spectrum}) and can be
trivially shown in exact analogy to the Schwarzschild case to fulfil:
\begin{equation}\label{gammaRN}
\alpha_{\pm}=\gamma_{\pm}.
\end{equation}
This implies together with Eq.(\ref{BekensteinRN}) the following
inequality: $\alpha_{+}>0>\alpha_{-}$.  (We will still use the
$\alpha_{\pm}$ notation to distinguish it from $\gamma_{\pm}$, but we
will use the Eq.(\ref{gammaRN}) in the following calculations.)

The coefficients $C_{1\pm}, C_{2\pm}, C_{3\pm}$ can be calculated as:
\[ C_{1\pm}=\frac{\sqrt{\pi}(1\pm\sqrt{1-K})}{\sqrt{1-K}},\]
\[C_{2\pm}=\pm\frac{2\sqrt{\pi}}{\gamma_{\pm}}\left(2-\frac{K(1+K)}{1\pm\sqrt{1-K}}\right),\]
\[C_{3\pm}=\pm\frac{2\sqrt{\pi}}{\gamma_{\pm}}\left(2-\frac{K}{1\pm\sqrt{1-K}}\right).\]
Assuming that $C_{2\pm}\alpha_{\pm}\mp 1>0$ and $C_{2+}+C_{3+}>0$ the
positivity of the occupation number condition leads to:
\begin{equation}\label{lowerRN}
f_{+}(0,\phi_{0\pm})>\frac{C_{1+}}{C_{2+}+C_{3+}}=\frac{\gamma_{+}(1+\sqrt{1-K})}{2\sqrt{1-K}\left(4-\frac{K(2+K)}{1+\sqrt{1-K}}\right)}>0,
\end{equation}
and
\begin{equation}\label{upperRN}
f_{-}(0,\phi_{0\pm})<\frac{C_{1-}}{C_{2-}+C_{3-}}=-\frac{\gamma_{-}(1-\sqrt{1-K})}{2\sqrt{1-K}\left(4-\frac{K(2+K)}{1-\sqrt{1-K}}\right)}.
\end{equation}

To determine the range of validity of the conditions $C_{2\pm}\mp 1>0$
and $C_{2\pm}+C_{3\pm}>0$ we needed to do some numerical calculations,
as the expressions contain polynomials of higher than second order.
Although these are in general analytically tractable, it is easier to
calculate and plot the functions numerically. Our numerical
calculations show the conditions $C_{2\pm}\alpha_{\pm}\mp 1>0$ and
$C_{2+}+C_{3+}>0$ are always true for the relevant $K$. However the
condition $C_{2-}+C_{3-}>0$ is violated for for $K\gtrapprox
0.85$. For this case the inequality given by Eq.(\ref{upperRN}) turns
to be the opposite and gives a lower bound.  For $K=0$, which is the
most relevant case:
\[C_{1+}=2\sqrt{\pi}, ~~~~C_{2+}=C_{3+}=\frac{4\sqrt{\pi}}{\gamma_{+}},\]
and therefore
\[f_{+}(0,\phi_{0})>\frac{\gamma_{+}}{4}.\] 
This is the Schwarzschild result from the previous section.  We also
see that the bounds on $f_{\pm}$ go to $\infty$ as we approximate the
extremal case (which is however outside the scope of our approach).

Now consider the first law of thermodynamics and calculate the
entropy:
\begin{equation}\label{FLTRN}
dS_{\pm}=\frac{dE_{\pm}}{T_{\pm}}-f_{\pm}(T_{\pm},\phi_{\pm})dN_{\pm}-\frac{\phi_{\pm}}{T_{\pm}}dQ.
\end{equation}
The first law of R-N black hole thermodynamics that comes from the
``correct'' entropy is:
\begin{equation}\label{FLToneRN}
dS_{\pm}=\pm\left(\frac{dM}{T_{\pm}}-\frac{\phi_{\pm}\cdot dQ}{T_{\pm}}\right)=2\pi M(1\pm\sqrt{1-K})^{2}dM.
\end{equation} 
The terms in Eq.(\ref{FLTRN}) read as:
\[\frac{d E_{\pm}}{T_{\pm}}=\pm\frac{2\pi(1+K)(1\pm\sqrt{1-K})^{2}M}{\sqrt{1-K}}dM,\]
\[dN_{\pm}=\frac{8\pi}{\gamma_{\pm}}\sqrt{1-K}(1\pm\sqrt{1-K})MdM, \]
furthermore
\[\frac{\phi_{\pm}}{T_{\pm}}=\pm\frac{4\pi}{\gamma_{\pm}}\frac{M\sqrt{1-K}\sqrt{K}}{(C_{2\pm}\alpha_{\pm}\mp  1)}\left\{f_{\pm}(\Delta_{\pm}\alpha_{\pm}+2)-C_{1\pm}\alpha_{\pm}\right\}, \]
with
\[\Delta_{\pm}=C_{3\pm}-C_{2\pm}=\pm\frac{2\sqrt{\pi}}{\gamma_{\pm}}\frac{K^{2}}{(1\pm\sqrt{1-K})}.\]

Then the first law of thermodynamics turns into:
\begin{eqnarray}\label{FLTtwoRN}
  dS_{\pm}=\left\{\pm\frac{  (1\pm\sqrt{1-K})}{\sqrt{1-K}}\left((1+K)(1\pm\sqrt{1-K})+\frac{2\sqrt{\pi}\sqrt{1-K}\cdot K}{C_{2\pm}\alpha_{\pm}\mp 1}\right)-\right.~~~~~\nonumber\\
  \left. f_{\pm}(0,\phi_{0\pm})\cdot \frac{2 \sqrt{1-K}}{\gamma_{\pm}}\left(2(1\pm\sqrt{1-K})\pm\frac{K(\Delta_{\pm}\alpha_{\pm}+2)}{C_{2\pm}\alpha_{\pm} \mp 1}\right)\right\}\cdot 2\pi M\cdot dM.~~
\end{eqnarray}
Now $f_{\pm}(0,\phi_{0\pm})$ is determined from equating the
expressions on the right side of Eq.(\ref{FLToneRN}) and
Eq.(\ref{FLTtwoRN}).  One immediate minimal condition is that the
entropy of the outer horizon grows with its area (the proportionality
is a positive number):
\begin{eqnarray}\label{cond1RN}
\frac{(1+\sqrt{1-K})}{\sqrt{1-K}}\left((1+K)(1+\sqrt{1-K})+\frac{2\sqrt{\pi}\sqrt{1-K}\cdot K}{C_{2+}\alpha_{+} - 1}\right)-~~~~~\nonumber\\
 f_{+}(0,\phi_{0+})\cdot \frac{2 \sqrt{1-K}}{\gamma_{+}}\left(2(1+\sqrt{1-K})+\frac{K(\Delta_{+}\alpha_{+}+2)}{C_{2+}\alpha_{+} - 1}\right)>0.
\end{eqnarray}

The same condition gives for the inner horizon:
\begin{eqnarray}\label{cond2RN}
  -\frac{  (1-\sqrt{1-K})}{\sqrt{1-K}}\left((1+K)(1-\sqrt{1-K})+\frac{2\sqrt{\pi}\sqrt{1-K}\cdot K}{C_{2-}\alpha_{-}+1}\right)-~~~~~\nonumber\\
  f_{-}(0,\phi_{0-})\cdot \frac{2 \sqrt{1-K}}{\gamma_{-}}\left(2(1-\sqrt{1-K})-\frac{K(\Delta_{-}\alpha_{-}+2)}{C_{2-}\alpha_{-}+1}\right)<0.
\end{eqnarray}
For $K=0$ the condition (\ref{cond1RN}) leads, as expected, to the
Schwarzschild result:
\[f_{+}(0,\phi_{0+})<\frac{\gamma_{+}}{2}.\] 
Also we can easily observe that the upper bound from Eq.(\ref{cond1RN}) for
$f_{+}(0,\phi_{0+})$ goes to $\infty$ as $K\to 1$ and the lower bound
for $f_{-}(0,\phi_{0-})$ from Eq.(\ref{cond2RN}) goes to
$-\infty$. This is expected and consistent with the condition for
positivity of occupation numbers for low $T_{+}$.

\begin{figure}[!htb]
 \centering
 \includegraphics[scale=.3, angle=270]{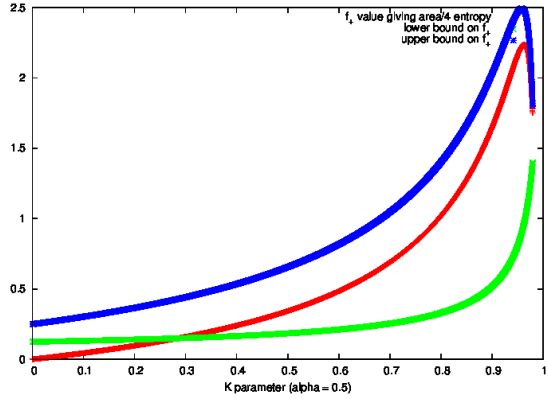}
 \caption{Plot showing the value of $f_{+}(0,\phi_{0+})$ for the correct proportionality of entropy and horizon area (red line), the lower bound on $f_{+}(0,\phi_{0+})$ (green line), and the upper bound on $f_{+}(0,\phi_{0+})$ (blue line). We see that the model is consistent (upper bound we get higher than lower bound).}
 \label{fig:rn}
 \end{figure}
 
 \begin{figure}[!htb]
  \centering
  \includegraphics[scale=.3, angle=270]{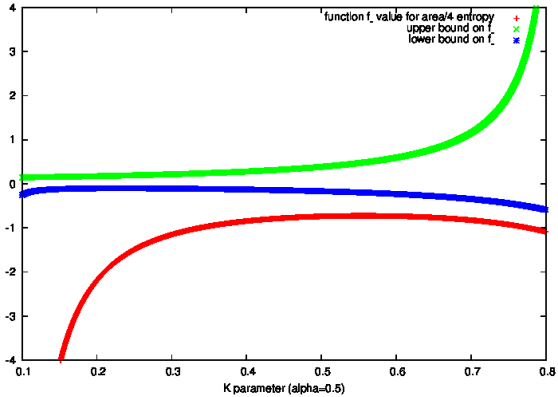}
  \caption{Plot showing the value of $f_{-}(0,\phi_{0-})$ for the correct proportionality of entropy and horizon area (red line), the lower bound on $f_{-}(0,\phi_{0-})$ (blue line), and the upper bound on $f_{-}(0,\phi_{0-})$ (green line). We see that the model is consistent (upper bound we get higher than lower bound).} 
    \label{fig:rnI}
  \end{figure}

  Numerical analysis of conditions (\ref{lowerRN}) and (\ref{cond1RN})
  for $f_{+}(0,\phi_{0+})$ is presented in Fig.(\ref{fig:rn}). (We
  used $\alpha_{+}=0.5$, but since all the results depend linearly on
  $\alpha_{\pm}$, a concrete choice of $\alpha_{\pm}>0$ is irrelevant
  for the analysis.) We see that the theory is consistent in the sense
  that lower bound is always smaller than the upper bound, therefore
  there always exist a range of choices of $f_{+}(0,\phi_{0+})$ for
  which the theory is sensible. (These choices guarantee no negative
  occupation numbers and also the increase of entropy as proportional
  to the area of the horizon.) This is very important to know. The red
  line in Fig.(\ref{fig:rn}) represents $f_{+}(0,\phi_{0+})$, such
  that leads to the correct proportionality between horizon area and
  entropy given by the factor $1/4$. As we see from
  Fig.(\ref{fig:rn}), the choice leading to $1/4$ is possible only for
  $K\gtrapprox 2.7$ and for low charges $K\lessapprox 2.7$ one cannot
  obtain the correct proportionality $S=A/4$. This is consistent with
  what we observed for the Schwarzschild case ($K=0$).

  The same investigation was done for the $f_{-}(0,\phi_{0-})$
  function and is presented in Fig.(\ref{fig:rnI}). The numerical
  results show that for $K\lessapprox 0.1$ we obtain from both
  conditions (\ref{upperRN}) and (\ref{cond2RN}) upper bounds on
  $f_{-}(0,\phi_{0-})$, therefore one can always fulfil these
  conditions. The same happens with lower bounds and $K\gtrapprox
  0.85$ (both of the conditions lead for such $K$ to lower
  bounds). The relevant regime where we needed to explore the
  consistency of the theory is $0.1\lessapprox K \lessapprox 0.85$. In
  this regime the Fig.(\ref{fig:rnI}) indeed shows that the theory is
  consistent, since the upper bound is always larger than the lower
  bound. The red line represents again the value of
  $f_{-}(0,\phi_{0-})$ for the correct proportionality between area
  and entropy ($1/4$ factor) and in this case one sees from
  Fig.(\ref{fig:rnI}) that the correct proportionality cannot be
  reached for any value of $K$.

\section{Discussion}

We have shown that equation of state for the holographic fluid living
on a spacetime horizon derived by \cite{Paddy1} is for stationary
configurations equal to an equation of state of an relativistic ideal
gas. This provides better physical understanding of the nature of the
fluid. Also the result is valid for \emph{generic} spacetime horizon
and depends only on General Relativity theory.

Many consequences of our microscopic holographic model for the black
hole thermodynamics were discussed in the appropriate places in the
text. The first main feature of our model is that it predicts for the
black hole horizon an equispaced area / entropy spectrum. (There are
many different arguments that lead to this type of spectrum that was originally
suggested by Bekenstein \cite{Bekenstein}. For overview of these
arguments see for example \cite{Skakala}.) The spacing in the area
spectrum is undetermined by our model, but it is equal to two times
ground state energy of a single particle state. One can then proceed
further and use either reduced quantization methods \cite{Kunstatter1,
  Kunstatter2}, or quasinormal modes \cite{Hod, Maggiore} to fix the
spacing between the area/entropy levels. The most widely agreed
outcome will give in Planck units the area spacing as $8\pi$ and will
lead to the one-particle's ground state energy equal to $4\pi T$.

The second prediction is that we recovered with our model the correct
type of proportionality between horizon area and entropy. However the
proportionality factor is for Schwarzschild black hole \emph{less
  than} one half of what is the result for black holes in Einstein
gravity. (We have also shown that the correct proportionality can be
reached in case of Reissner-Nordstr\"om black hole for sufficiently
large $|Q|/M$ ratio.) One can try to ``resolve'' this problem by
suggesting more species of particles whose entropy adds up to the
correct proportionality, but this might sound slightly artificial
without support of some additional arguments. Anyway, let us keep such
a suggestion open for future investigations.

It is very important to realize some limitations of our approach:
Standardly one can fully derive macroscopic fluid dynamics from a
Hamiltonian of a microscopic theory. However despite of
instructiveness of our model, this is not going to be possible with
our microscopic model, at least not in the usual sense. One does not
posses a local concept of energy in General Relativity in dynamical
situations and this could pose a fundamental problem in deriving some
local fluid dynamics from our (and any other similar) model. It is
also clear that the model of free fluid would be unable to explain the
dissipative terms that appear in the equations derived by Damour
\cite{Damour} and contribute significantly to the dynamics. So even if
the energy problem was overcome one would expect that the ``full''
microscopic model will have some more complicated features and only in
the limit of stationarity the EQS reduces to a free relativistic
gas. To get an improved microscopic model from which full dynamics
could be derived is an open problem.

There are many more open questions and problems, for example:
\begin{itemize}
\item[a)] We considered only the simplest, scalar (spin 0) Bose
  particle model. One does not assume any significant changes for
  bosons if higher spins $s=1,2$ were employed, but one needs to
  calculate the one-particle energy spectra for such cases. Suggesting
  that the spectra should be similar to the spectrum (\ref{spectrum}),
  one can also exclude half spin particles (fermions). This
  calculation is done in Appendix \ref{A}. However all this needs to
  be further explored.
\item[b)] Only the simplest models of Schwarzschild and
  Reissner-Nordstr\"om black holes were explored, it would be nice to
  see to what extent we can generalize the model to the stationary
  rotating (Kerr) black hole.
\item[c)] Many results \cite{Paddy2, Paddy3, Skakala} suggest that
  distinguishing between black hole horizon and more general spacetime
  horizons is somewhat artificial. To what extent can our results be
  generalized beyond black holes? Also, since our results are for
  fundamental reasons non-local (energy spectra are a non-local
  quantity), is there still a chance to somehow reformulate the theory
  for a local Rindler horizon?
\item[d)] Another question are generalized theories of gravity. Is a
  similar microscopic model applicable to black holes / horizons in
  generalized theories of gravity? The fact that in many generalized
  theories entropy follows Bekenstein quantization rules, but the
  horizon area does not \cite{Sarkar}, suggests that generalization
  within gravity theories might correspond to a generalization within
  fluid's equation of state. Is this indeed true? Can one classify
  gravity theories by the equations of states of the fluid? The most
  straightforward generalization of Einstein equations, by introducing
  cosmological constant $\Lambda$, suggests that the answer is
  positive. It can be easily shown that the fluid's equation of state
  acquires for the theory with $\Lambda$ an additional term
  proportional to $\sqrt{V}$:
  \[p=\frac{E}{2V} - \frac{\Lambda \sqrt{V}}{6(4\pi)^{3/2}}.\] Also some
  additional questions relating to what was previously said can be
  raised: For example, can one interpret the non-equilibrium
  thermodynamical effects that occur in higher curvature gravity
  \cite{Jacobson2, Liberati2} through such a microscopic model of a
  suitable fluid?
\item[e)] One can also stick to the model we have and further explore,
  and discuss, possible consequences of the model, such that could
  lead to potentially observable predictions.
\end{itemize}

\medskip

\section*{Acknowledgments}

We would like to thank T. Padmanabhan for discussions. The work is
supported by Max Planck-India Partner Group on Gravity and
Cosmology. SS is partially supported by Ramanujan Fellowship of DST,
India.

\begin{appendix}

\section{Fermionic model with energy spectrum (\ref{spectrum2})\label{A}}

Let us consider Schwarzschild black hole and fermions assuming that
the one-particle energy spectrum will be the same as in case of zero
spin particles, hence the spectrum is given by
Eq.(\ref{spectrum2}). [There will be small corrections due to the spin
1/2 nature, but in the present calculation they are not supposed to
contribute.]

Let us first estimate the function $N(E)$ for temperatures so close to
zero, that all the energy states up to Fermi level will be occupied by
one particle, and the levels above will have zero occupation
number. The energy can be calculated as:
\begin{equation}\label{fermi en}
E=\sum_{\ell=0}^{L}g_{\ell}\cdot\epsilon_{\ell}=\sum_{\ell=0}^{L}(2\ell+1)\cdot \ell T.
\end{equation}
(Here we approximated the spectrum (\ref{spectrum2}) for higher levels
by $\epsilon_{\ell}\approx T\ell $ and $L$ labels the Fermi level.)
The number of particles $N$ is given as:
\begin{equation}\label{Fermi N}
N=\sum_{\ell=0}^{L}(2\ell+1).
\end{equation}

Calculating the sum in Eq.(\ref{Fermi N}) one obtains:
\[N=L^{2}+2L.\]
Considering that $N>>1$ one further obtains:
\[L\approx\sqrt{N}.\]
One can also calculate the sum in Eq.(\ref{fermi en}) which leads to:
\[E=\left(\frac{L(L+1)(2L+1)}{3}+L\right)\cdot T.\]
The dominant term gives:
\[E\approx\frac{2}{3}N^{3/2}T,\]
and then using our model constraints
\[N=(12\pi)^{2/3}E^{4/3}.\]

The state's occupation number can be similarly to the bosonic case
derived as:

\begin{equation}
\label{Fermi n}
\bar{n}_{\ell}=\frac{g_{\ell}}{\exp\left(\left\{\frac{1}{T}+\frac{p}{T}64\pi E 
      -\frac{4}{3}(12\pi)^{2/3}\left(\frac{4}{3}\frac{\mu}{T}-\lambda\right)E^{1/3}\right\}\epsilon_{\ell}-\lambda\right)+1}.
\end{equation}

This with our constraints leads to the result:

\begin{equation}
\label{Fermi n2}
\bar{n}_{\ell}=\frac{g_{\ell}}{\exp\left(\left\{3-\frac{4}{3}\left(\frac{18\pi}{T}\right)^{1/3}\left(\frac{4}{3}\mu-\lambda T\right)\right\}\tilde\epsilon_{\ell}-\lambda\right)+1}.
\end{equation}

Now the first law of thermodynamics leads to:
\begin{equation}\label{first fermi}
dS=\frac{\left(2-\frac{\mu}{T^{1/3}}(18\pi)^{1/3}\frac{4}{3}\right)dE}{T}.
\end{equation}
To have the correct asymptotic proportionality between area and
entropy $\mu$ has to fulfill:
\[\mu=\mu_{0}T^{\frac{1}{3}}+O(T^{\alpha}),~~~~~~\alpha>\frac{1}{3}.\]

The infinite number of particles in the asymptotic $T=0$ case can be
obtained from Eq.(\ref{Fermi n}) when:
\[\mu_{0}\geq \frac{27}{16 (18\pi)^{1/3}}.\]
But this yields:
\[2-\mu_{0}(18\pi)^{\frac{1}{3}}\frac{4}{3}\leq 2-\frac{9}{4}=-\frac{1}{4},\]
which gives from Eq.(\ref{first fermi}) entropy decreasing with
energy, instead of growing with energy. This argument excludes Fermi
particles.

\end{appendix}

\end{document}